Please make sure you insert your
\documentclass[a4paper,11pt]{article}
\usepackage{pos}
\usepackage{graphicx}
\usepackage{tikz}
\usepackage{subcaption}
\usepackage{enumitem}
\usepackage{verbatim}
\usepackage{titlesec}
\titlespacing\section{0pt}{10pt}{10pt}

\newcommand{\e}{\vec e}
\newcommand{\logchi}{6.2 }
\newcommand{\powchi}{5.5 }
\graphicspath{{figures/}}

\let\OLDthebibliography\thebibliography
\renewcommand\thebibliography[1]{
  \OLDthebibliography{#1}
  \setlength{\parskip}{0pt}
  \setlength{\itemsep}{0pt plus 0.1ex}
}

\title{Topology of the $O(3)$ non-linear sigma model under the gradient flow}

\author*[a]{Stuart Yi-Thomas}
\author[a,b]{Christopher Monahan}

\affiliation[a]{Department of Physics, William \& Mary, \\ 300 Ukrop Way, Williamsburg, VA, USA} 
\affiliation[b]{Thomas Jefferson National Accelerator Facility, \\ 12000 Jefferson Avenue, Newport News, VA, USA} 

\emailAdd{snthomas01@email.wm.edu}
\emailAdd{cjmonahan@wm.edu}

\abstract{
The $O(3)$ non-linear sigma model (NLSM) is a prototypical field theory for QCD and ferromagnetism, and provides a simple system in which to study topological effects. In lattice QCD, the gradient flow has been demonstrated to remove ultraviolet singularities from the topological susceptibility. In contrast, lattice simulations of the NLSM find that the topological susceptibility diverges in the continuum limit, even in the presence of the gradient flow. We introduce a $\theta$-term and analyze the topological charge as a function of $\theta$ under the gradient flow. Our results show that divergence persists in the presence of the flow, even at non-zero $\theta$.
}

\FullConference{%
 The 38th International Symposium on Lattice Field Theory, LATTICE2021
  26th-30th July, 2021
  Zoom/Gather@Massachusetts Institute of Technology
}

\begin{document}
\maketitle

\section{Introduction}
\label{sec:introduction}
Spin models provide a framework for understanding the physics of strongly-coupled systems, from solid state and condensed matter systems to nuclear and particle physics. The non-linear sigma model (NLSM), in particular, has provided a rich arena in which to study nonperturbative effects. In solid-state systems, this model describes Heisenberg ferromagnets \cite{callan1985} and in nuclear physics, it acts as a prototype for quantum chromodynamics (QCD), the gauge theory of the strong nuclear force. In general, the NLSM shares key features with non-Abelian gauge theories such as QCD, including a mass gap and asymptotic freedom \cite{polyakov1975}, and has proved a useful model for exploring the effect of these properties in simple systems.

We consider the $O(3)$ NLSM in 1+1 dimensions (one dimension of space, one dimension of time). This theory may exhibit topological effects, such as \textit{instantons}, which are classical field solutions at local minima of the action in Euclidean space. These topologically protected solutions cannot evolve into the vacuum state via local fluctuations. This property has become critically important to understanding several applications of quantum field theories in cosmology and high energy physics, such as the existence of magnetic monopoles \cite{goddard1986} and the mass of the $\eta'$ Goldstone boson \cite{witten1979, veneziano1979}. Additionally, topological stability may become a key tool for fault-tolerant quantum computers \cite{kitaev1997}. In these devices, topology protects the delicate quantum states necessary for information processing.

The protection of topological instantons in the 1+1 $O(3)$ NLSM relies on a vanishing topological susceptibility. However, the convergence properties of this quantity are still unclear \cite{bietenholz2018,Berni:2020ebn}. Analytical arguments suggest the topological susceptibility should approach zero in the continuum limit \cite{berg1981}, but numerical results on the lattice, summarized in \cite{bietenholz2018}, indicate a logarithmic divergence and support the semi-classical picture of small-size instantons generating the divergence. Similar, but ultimately inconclusive, results were found for the equivalent CP$^{1}$ model in the presence of an alternative smoothing procedure, ``cooling'', in \cite{Berni:2020ebn}. To elucidate this apparent contradiction, we apply the gradient flow, a local smearing of operators which preserves gauge invariance. In quantum chromodynamics, this technique has corroborated a previous analytical result \cite{giusti2004} by removing ultraviolet divergences on the lattice \cite{bruno2014}. This success has motivated the gradient flow to as a tool to define the continuum limit of the topological susceptibility in the 1+1 $O(3)$ NLSM. Despite this intuition, recent studies demonstrate that the topological susceptibility still diverges in the continuum limit in the presence of the ultraviolet smearing procedures \cite{bietenholz2018,Berni:2020ebn}.

We also study a second perspective on the topological susceptibility arising from the introduction of a $\theta$-term into the field Lagrangian. This term drives the vacuum state into a topological phase \cite{allessalom2008}, and the resulting theory may exhibit confining, walking, and conformal behaviors at different values of $\theta$ \cite{Nogradi:2012dj}. Differentiating the partition function with respect to $\theta$ yields a value proportional to the topological susceptibility. The effect of nonzero $\theta$ on the theory therefore should reflect the divergence of the susceptibility in the continuum limit. In this work we verify the divergence of the topological susceptibility and develop a clearer picture of how the $\theta$-term affects the topology of the 1+1 $O(3)$ NLSM.

To study the topological qualities of the NLSM numerically, we first implement a Markov chain Monte Carlo simulation using Metropolis and Wolff cluster \cite{wolff1989} algorithms. The gradient flow has no exact solution in the NLSM, so we implement a numerical solution using a fourth-order Runge-Kutta approximation with automatic step sizing. 

\section{The non-linear sigma model}
\label{sec:nlsm}

We study the $O(3)$ NLSM in two dimensions, defined by the Euclidean action
\begin{equation*}
    \label{eq:nlsm euclidean action}
    S_E = \frac{\beta}{2} \int d^2x \; \sum_{i=1}^{2}\left(\partial_i \e\, \right)^2,
\end{equation*}
where $\e$ is 3-component real vector constrained by $|\e\,|=1$ and $\beta$ is the inverse coupling constant. Following~\cite{berg1981}, we define the topological charge density, $q(x^*)$, for each plaquette $x^*$ such that the total topological charge is
\begin{equation}
    Q = \sum_{x^*} q(x^*),
\end{equation}
where
\begin{equation}
  \label{eq:topodensity}
    q(x^*) = \frac{1}{4\pi} \bigg[A\Big(\e(x_1), \e(x_2), \e(x_3)\Big) + A\Big(\e(x_1), \e(x_3), \e(x_4)\Big) \bigg].
\end{equation}
Here, $A$ is the signed area of the triangle in target space. Visually, the three points in each of the two terms in Eq.~\ref{eq:topodensity} form halves of the plaquette, which we represent in Fig.~\ref{fig:plaquette}. The resultant signed area is represented in Fig.~\ref{fig:signedarea}. The value $A$ is defined if $A\neq 0, 2\pi$, or in other words, as long as the three points on the sphere are distinct and do not form a hemisphere. In numerical calculations, these points can be ignored. Therefore, we impose that the signed area is defined on the smallest spherical triangle, or $-2\pi < A < 2\pi$.
\begin{figure}
  \centering
    \begin{subfigure}[b]{0.4\textwidth}
    \centering
    \begin{tikzpicture}[scale=0.4, every node/.style={scale=0.9}]
        \draw[step=4cm,gray,thin] (-1,-1) grid (5,5);
        \draw[dotted] (0,0) -- (4,4);
        \draw[thick](0,0) node[circle,fill,inner sep=0pt, minimum size=0.2cm]{} node[anchor=north west] {$x_1$} --
                    (0,4) node[circle,fill,inner sep=0pt, minimum size=0.2cm]{} node[anchor=north west] {$x_2$} --
                    (4,4) node[circle,fill,inner sep=0pt, minimum size=0.2cm]{} node[anchor=north west] {$x_3$} --
                    (4,0) node[circle,fill,inner sep=0pt, minimum size=0.2cm]{} node[anchor=north west] {$x_4$} -- cycle ;
                    \draw[] (2,2) node[circle,fill,inner sep=0pt, minimum size=0.2cm]{} node[anchor=north west]{$x^*$};

        \draw[-stealth] (0.2,1) -- (0.2,3);
        \draw[-stealth] (1,3.8) -- (3,3.8);
        \draw[-stealth] (3,3.5) -- (1,1.5);

        \draw[-stealth] (1,0.5) -- (3,2.5);
        \draw[-stealth] (3.8,3) -- (3.8,1);
        \draw[-stealth] (3,0.2) -- (1,0.2);
    \end{tikzpicture}
    \caption{\label{fig:plaquette} Visualization of a plaquette $x^*$. The dotted line separates the plaquette into two signed areas, used to define the topological charge density $q(x^*)$. Arrows represent the order of signed area.}
    \vspace*{-\baselineskip}
    \end{subfigure}
    \hspace*{1cm}
    \begin{subfigure}[b]{0.4\textwidth}
    \centering
    \begin{tikzpicture}[scale=0.6, every node/.style={scale=1}]
        \pgfdeclarelayer{nodelayer}
        \pgfdeclarelayer{edgelayer}
        \pgfdeclareradialshading{sphere4}{\pgfpoint{-0.2cm}{0.5cm}}%
            {rgb(0cm)=(1,1,1);
            rgb(1cm)=(0.5,0.5,0.5); rgb(1.05cm)=(1,1,1)}
        \pgfsetlayers{nodelayer,edgelayer}
        \tikzstyle{label}=[fill=none, draw=none, shape=circle]
        \tikzstyle{point}=[inner sep=0pt, minimum size=0.2cm,fill=black, draw, shape=circle]
        \tikzstyle{interior line}=[dotted,thick]

        \tikzstyle{triangle}=[thick]
        \tikzstyle{arrow}=[->, thick]
        \begin{pgfonlayer}{nodelayer}
            \node [style=point] (4) at (0.5, 2.5) {};
            \node [style=point] (5) at (0.55, 0.625) {};
            \node [style=point] (6) at (2.25, 0.75) {};
            \node (7) at (0, 0) {};
            \node (8) at (1.25, 1.375) {};
            \node (9) at (2.425, 2.525) {};
        \end{pgfonlayer}
        \begin{pgfonlayer}{edgelayer}
            \draw (0,0) circle (3cm);
            \shade[shading=sphere4] (0,0) circle (3cm);
            \draw [bend left=15,style=triangle] (4.center) to (6.center);
            \draw [bend left=355,style=triangle] (6.center) to (5.center);
            \draw [bend right=5,style=triangle] (5.center) to (4.center);
            \draw [style=interior line] (4.center) to (7.center);
            \draw [style=interior line] (5.center) to (7.center);
            \draw [style=interior line] (6.center) to (7.center);
            \draw [fill=gray!80] (4.center) to [bend left=15] (6.center) to [bend left=355] (5.center) to [bend right=5] cycle;


            \node [style=label, anchor=north] at (2.25, 0.75) {\small $\e(x_1)$};
            \node [style=label, anchor=north] at (0.55, 0.625) {\small $\e(x_2)$};
            \node [style=label, anchor=east] at (0.5, 2.5) {\small $\e(x_3)$};
            \node [style=point] at (4){};
            \node [style=point] at (5){};
            \node [style=point] at (6){};

            \node [style=label] at (8) {\small $A$};
        \end{pgfonlayer}
    \end{tikzpicture}
    \caption{\label{fig:signedarea} The signed area of a triangle in target space.}
    \end{subfigure}
    \caption{\label{fig:topology}}
\end{figure}

From the definition of $Q$, we define the topological susceptibility as
\begin{equation}
\chi_t \equiv \frac{1}{L^2} \Big( \langle Q^2 \rangle - \langle Q \rangle^2 \Big).
\end{equation}
Since $\langle Q \rangle=0$ in the NLSM, this quantity becomes
\begin{equation}
    \chi_t = \frac{1}{L^2} \sum_{x^*} \langle q(x^*)q(0)\rangle,
\end{equation}
where we have assumed periodic boundary conditions. In the continuum limit, the topological susceptibility diverges owing solely to the $x^*=0$ term \cite{bietenholz2018}, a divergence that also exists in QCD \cite{bruno2014}. In the semiclassical approximation, this divergence arises from small-size instantons \cite{Luscher:1981tq}.

As an extension, we can generalize the NLSM to have a nonzero vacuum expectation value for the topological charge ($\langle Q \rangle\neq0$), manifested by the addition of a ``$\theta$-term'':
\begin{equation*}
    S[\e\,] \rightarrow S[\e\,] - i \theta Q[\e\,].
\end{equation*}
defining the ``nontrivial'' NLSM. We find that the topological susceptibility in the trivial case depends on the charge in the nontrival model:
\begin{equation}
    \chi_t \propto \left. \frac{d\,\mathrm{Im}\langle Q \rangle}{d\theta}\right|_{\theta=0}
\end{equation}
This final relation will allow us to probe the behavior of the topological susceptibility in the trivial NLSM by considering the topological charge in the nontrivial theory.

\section{Gradient flow}
\label{sec:gradflow}
To resolve the ultraviolet divergence in the $\chi_{t}$, we adopt a technique known as ``smearing'', a local averaging of the field \cite{solbrig2008}. Specifically, we use the gradient flow \cite{,Narayanan:2006rf,Luscher:2010iy,Luscher:2010we,monahan2015,monahan2016}, which introduces a new half-dimension\footnote{The term ``half-dimension'' indicates that the flow time is exclusively positive.} called ``flow time''.  The flow time $\tau$ parameterizes the smearing such that an evolution in flow time corresponds to suppressing ultraviolet divergences, pushing field configurations toward classical minima of the action.

We can choose any flow time equation that drives the field towards a classical minimum. Following \cite{Makino:2014sta,Kikuchi:2014rla,bietenholz2018}, we can define the gradient flow for the NLSM via the differential equation
\begin{equation}
    \label{eq:nlsm_gradflow}
  \partial_\tau \e (\tau,x) = \left( 1 - \e(\tau,x) \e(\tau,x)^T \right) \partial^2 \e(\tau,x),
\end{equation}
which we solve numerically with the boundary condition $\e(\tau=0,x) = \e(x)$.

\section{Numerical implementation}
\label{sec:numerics}
We implement a numerical Monte Carlo method to study the NLSM in two dimensions using the discretized action

\begin{equation}
    \label{eq:nlsm discretized action}
    S_\mathrm{lat}[\e\,] = \sum_i \left[ 2 - \sum_{\mu=0}^{2}\e(x+a\hat{\mu}\,)\cdot\e(x)  \right]
\end{equation}
where $a$ is a lattice constant and $\hat \mu$ are the Euclidean unit vectors. We generate configurations using the Metropolis \cite{metropolis1953} and Wolff cluster \cite{wolff1989} algorithms.

\begin{figure}[h]
  \centering
      \begin{subfigure}[b]{0.5\textwidth}\centering
        \includegraphics[width=0.9\textwidth]{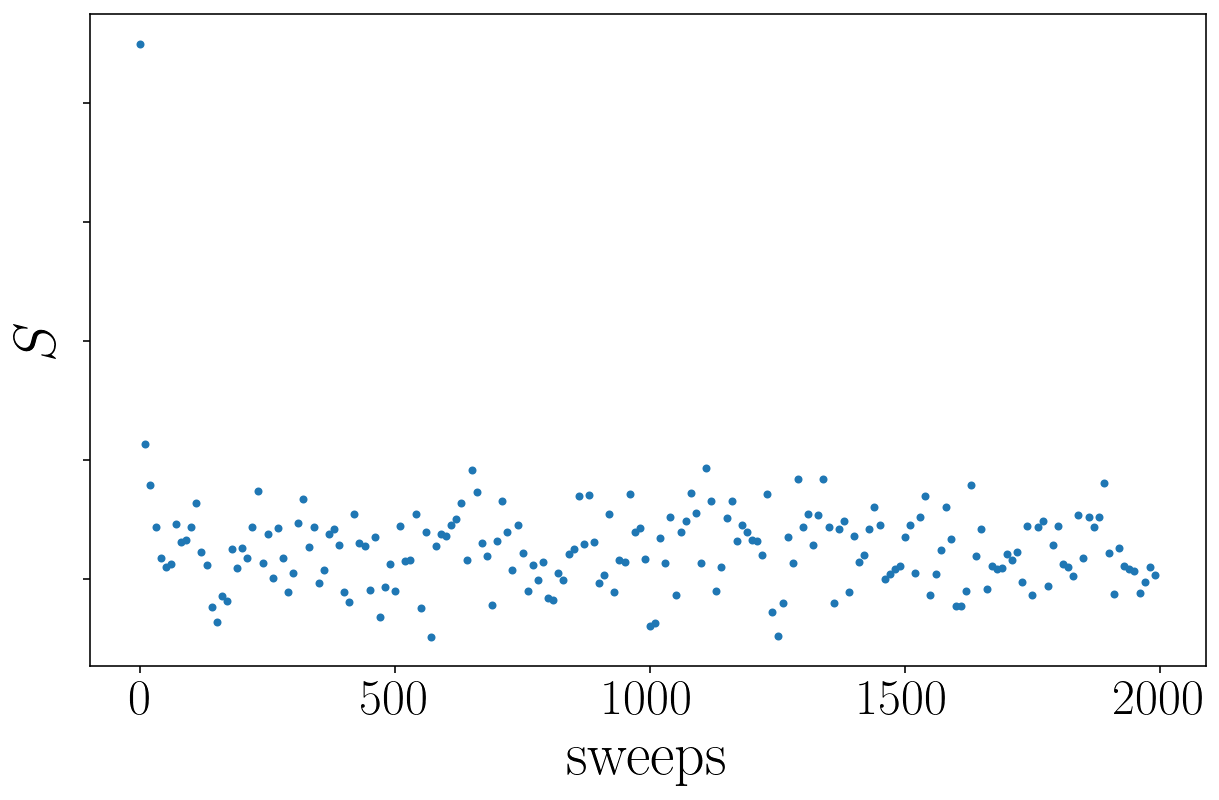}
        \caption{$L=24$}
      \end{subfigure}%
      \hfill
      \begin{subfigure}[b]{0.5\textwidth}\centering
        \includegraphics[width=0.9\textwidth]{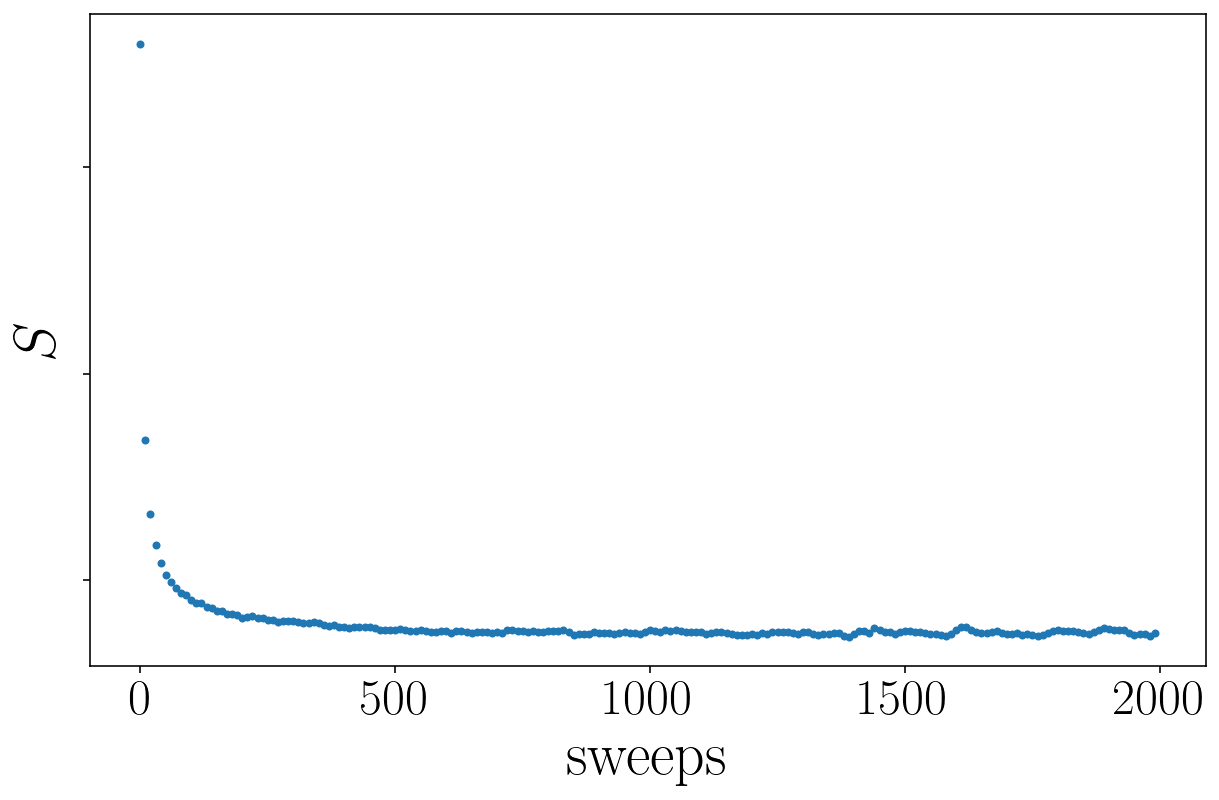}
        \caption{$L=404$}
      \end{subfigure}
      \hfill
      \caption{\label{fig:therm} Plots of the action, $S$, as a function of Monte Carlo time, starting with a random NLSM lattice.}
\end{figure}

We thermalize the configurations with 1000 sweeps, with a cluster update every five sweeps, and illustrate a sample Markov chain in Fig.~\ref{fig:therm}, where we plot the action as a function of Metropolis sweeps.
We use Wolff's automatic windowing procedure \cite{wolff2007} to estimate the autocorrelation times for various observables, such as the magnetic susceptibility $\chi_m$. We measure observables every $50$ sweeps for each simulation.

We apply the gradient flow equation (Eq.~\ref{eq:nlsm_gradflow}) replacing the continuous Laplacian operator $\partial^2$ with a discrete analogue
\begin{equation*}
    \partial^2 \e(\tau,x) = \e(\tau, x+a \hat{t}) + \e(\tau,x-a\hat t) + \e(\tau, x+a \hat{x}) + \e(\tau,x-a\hat x) - 4 \e(t,x).
\end{equation*}
and numerically solving the differential equation using a fourth-order Runga-Kutta approximation. To increase the efficiency of this algorithm, we implement the step-doubling algorithm to adaptively adjust the step size. If the error of a Runge-Kutta step is greater than the tolerance, the same step is repeated with half the step size. Alternatively, if the error is less than half of the tolerance, the step size is doubled for the next calculation. Finally, if the step size is greater than the distance to the next measurement, that distance is used as the step size, using the normal value afterwards. Otherwise, the algorithm proceeds with the consistent step size.

To calculate the error, we compare one lattice $\e_1$ produced using a step of size $2h$ with another lattice $\e_2$ produced via two steps of size $h$. The error $\Delta$ can be estimated to up the fifth order of $h$ as \cite{vetterling1992}
\begin{equation}
\Delta = \frac{1}{15}\sqrt{\sum_x \left| \e_2(x) - \e_1(x) \right|}
\end{equation}
The tolerance used in this work is $\Delta_{max}=0.01$.

\section{Results}
\label{sec:results}

\begin{figure}[h!]
    \centering
      \includegraphics[width=0.65\textwidth]{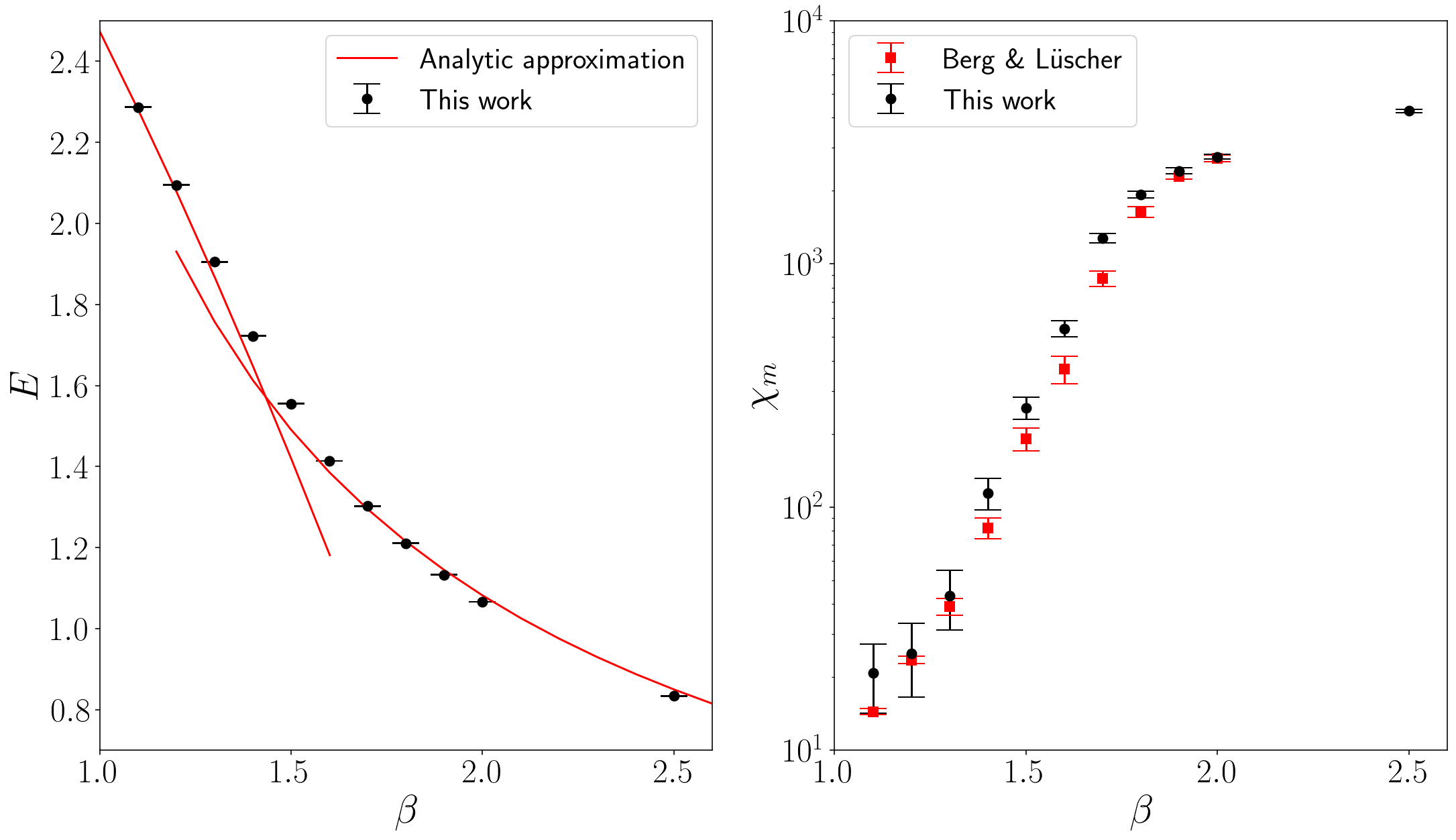}
      \caption{\label{fig:bergluscher} Comparison with \cite{berg1981}. First panel: internal energy compared with analytic  energy (Eq.~\ref{eq:analytic_energy}). Second panel: magnetic susceptibility compared with literature values.}
\end{figure}

We check our numerical implementation of the NLSM by comparison to studies of the internal energy and magnetic susceptibility in Refs.~\cite{berg1981} and \cite{bietenholz2018}. Following \cite{berg1981}, we approximate the internal energy in the strong ($\beta<1$) and weak ($\beta>2$) regimes as
\begin{equation}
    \label{eq:analytic_energy}
E \approx \begin{cases}
    4-4y-8y^3-\frac{48}{5}y^5& \beta<1 \\
\frac{2}{\beta} + \frac{4}{\beta^2} + 0.156\frac{1}{\beta^3}& \beta>2,
\end{cases}
\end{equation}
where $y=\operatorname{coth}\,\beta-1/\beta$.
We compare this analytical result and simulated values of $\chi_m$ with the Monte Carlo simulation in Fig.~\ref{fig:bergluscher}.
The slight discrepancy between numerical results in the cross-over region provides an estimate of unquantified systematic uncertainties.

We also seek to confirm the results from Bietenholz et al. \cite{bietenholz2018}. In addition, we show the topological susceptibility $\chi_t$ diverges in the continuum limit even at finite flow time. Since $\chi_t$ is in units of inverse distance squared, we multiply by , to achieve a scale-invariant value $\chi_t\xi_2^2$. Additionally, we use a parameter $t_0$ to scale the flow time such that $t_0\sim L^2$. In our Monte Carlo simulation, we utilize the same values as \cite{bietenholz2018} for $\xi_2$, $\beta$ and $t_0$.

\begin{figure}[h!]
    \centering
      \includegraphics[width=0.7\textwidth]{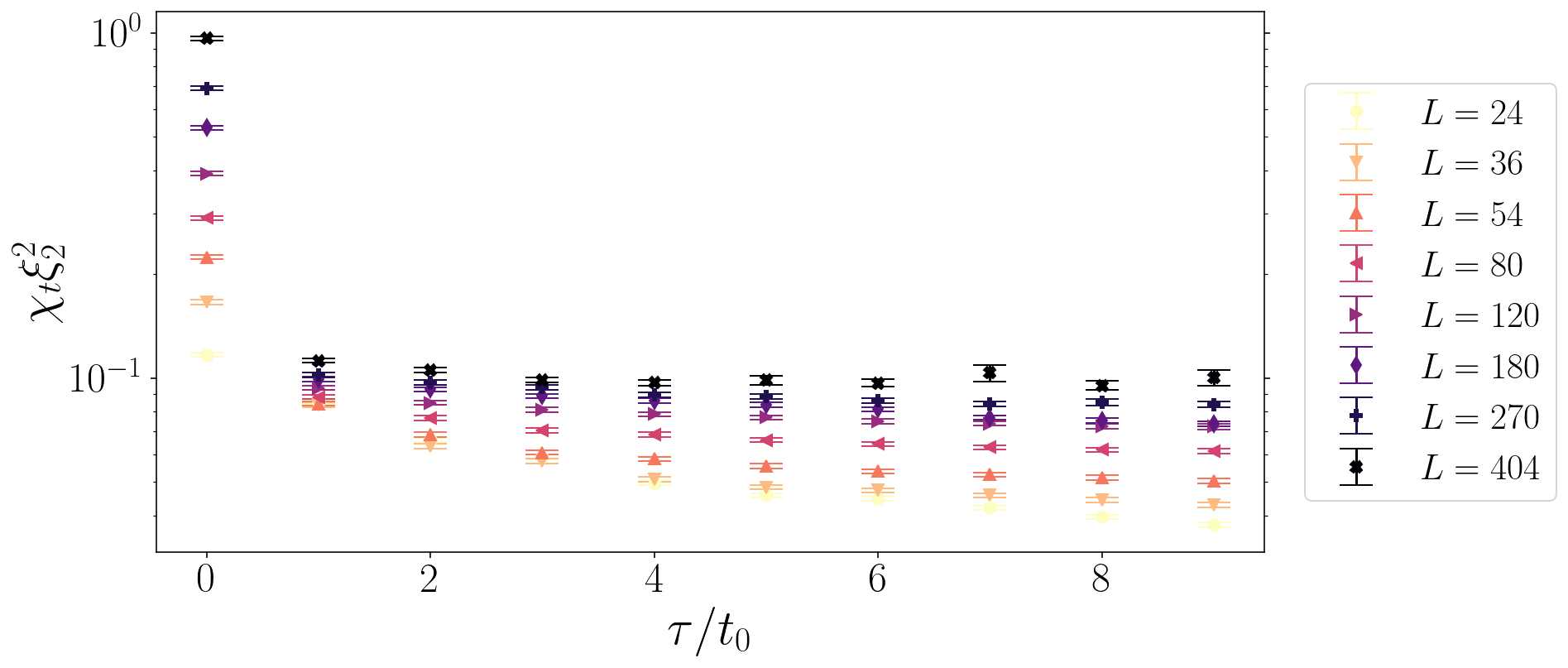}
      \caption{\label{fig:bietenholz} Dimensionless magnetic susceptibility, $\chi_t \xi_{2}^2$, as a  function of flow time $\tau/t_0$. Simulation run with 10,000 measurements every 50 sweeps, 1,000 sweep thermalization. }
\end{figure}

\begin{figure}[h!]
    \begin{center}
      \begin{subfigure}[b]{0.45\textwidth}
          \centering
          \includegraphics[height=0.7\textwidth]{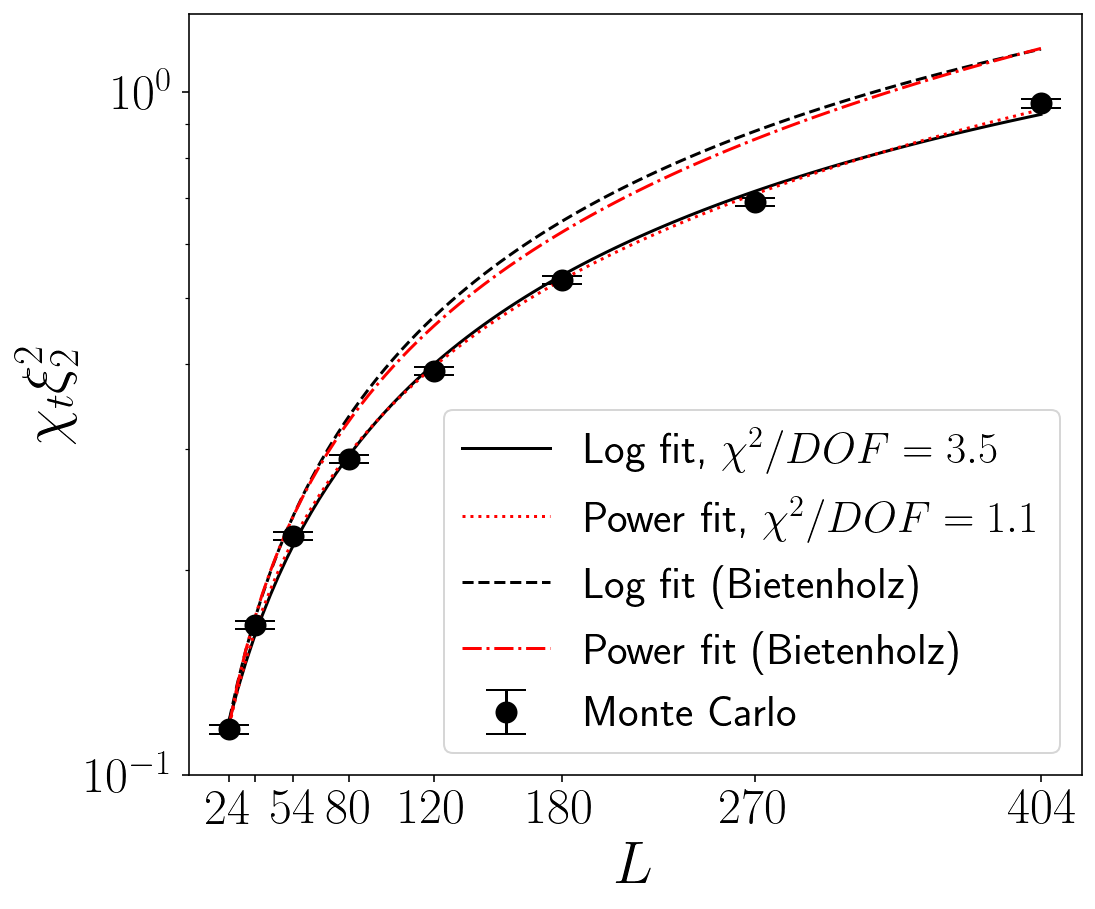}
          \caption{$\tau = 0$}
      \end{subfigure} %
      \begin{subfigure}[b]{0.45\textwidth}
          \centering
          \includegraphics[height=0.7\textwidth]{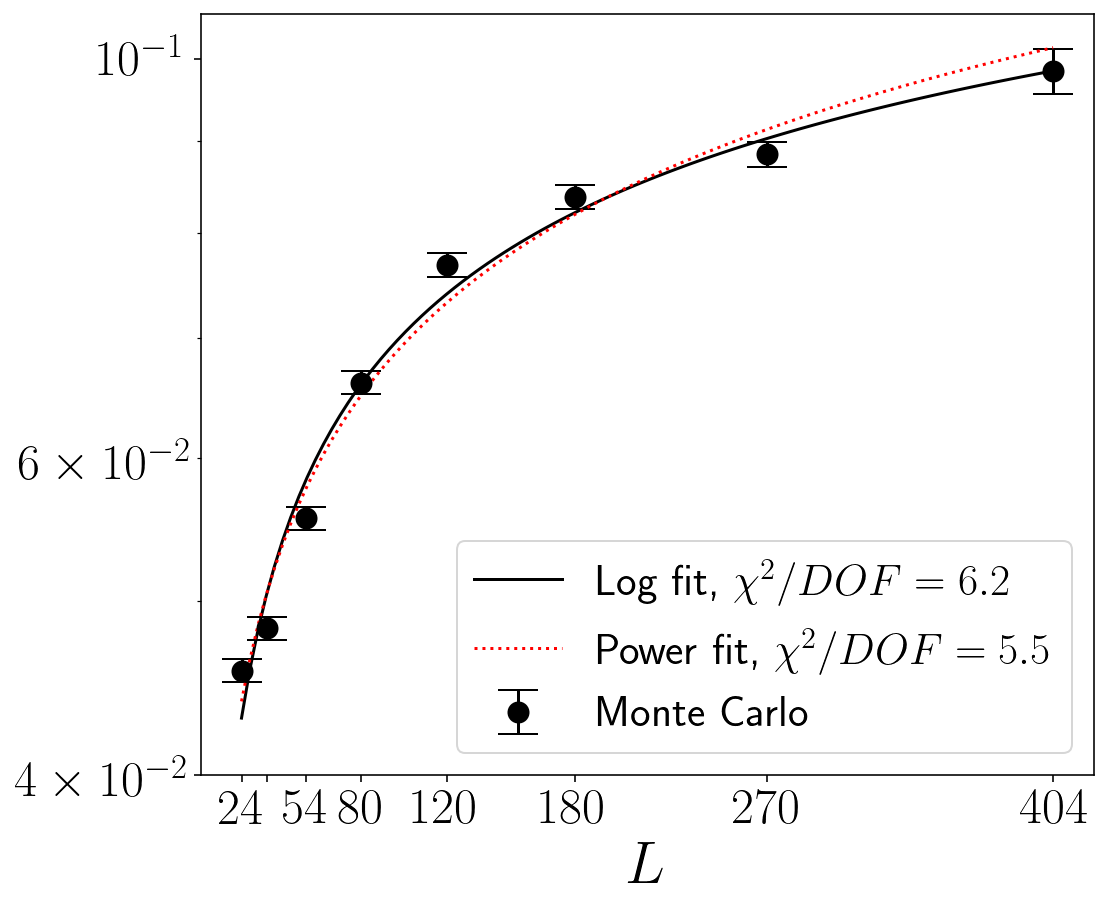}
          \caption{$\tau = 5t_0$}
      \end{subfigure}
      \caption{\label{fig:divergence} $\chi_t\xi_2^2$ as a function of $L$. We fit the data with both a logarithmic and power fit. Simulation run with 10,000 measurements, once every 50 sweeps, 1,000 sweep thermalization. In the $\tau=0$ case, we have compared our result with the curve fit found in \cite{bietenholz2018}.}
    \end{center}
\end{figure}

In Fig.~\ref{fig:bietenholz} we plot the dimensionless variable $\chi_t\xi_2^2$, where $\xi_2^2$ is the square of the second moment correlation length, as a function of flow time $\tau/t_0$, with $t_0 \sim L^2$.
Fig.~\ref{fig:bietenholz} shows that the flow time effectively decreases the topological susceptibility by dampening high-momentum modes. To analyze the divergence of $\chi_t$ in the continuum limit, we plot $\chi_t \xi_2^2$ as a function of lattice size $L$, for $\tau=0$ and $\tau=5t_0$, in Fig.~\ref{fig:divergence}.
We fit the data with two fit functions: a log fit
\begin{equation}
    \chi_t \xi_2^2 = a \mathrm{log}(b L + c);
\end{equation}
and a power law fit
\begin{equation}
    \chi_t \xi_2^2 = a L^b + c.
\end{equation}
We calculate the parameters to these functions using two independent fitting methods, the \texttt{lsqfit} package \cite{lsqfit} and the \texttt{scipy} \texttt{curve\_fit} tool \cite{virtanen2020}. We quote results from \cite{lsqfit}. For flow time $\tau=5t_0$, the $\chi^2/DOF$  for the logarithmic and power fits are \logchi and \powchi respectively, which potentially indicates underestimated systematic uncertainties. Both of these functions diverge as $L\rightarrow \infty$, indicating that the topological susceptibility also diverges in the continuum limit, in line with the results of \cite{bietenholz2018}. The origin of the discrepancies between our results and \cite{bietenholz2018} will be studied in future work.

\begin{figure}[h!]
    \centering
      \includegraphics[width=0.5\textwidth]{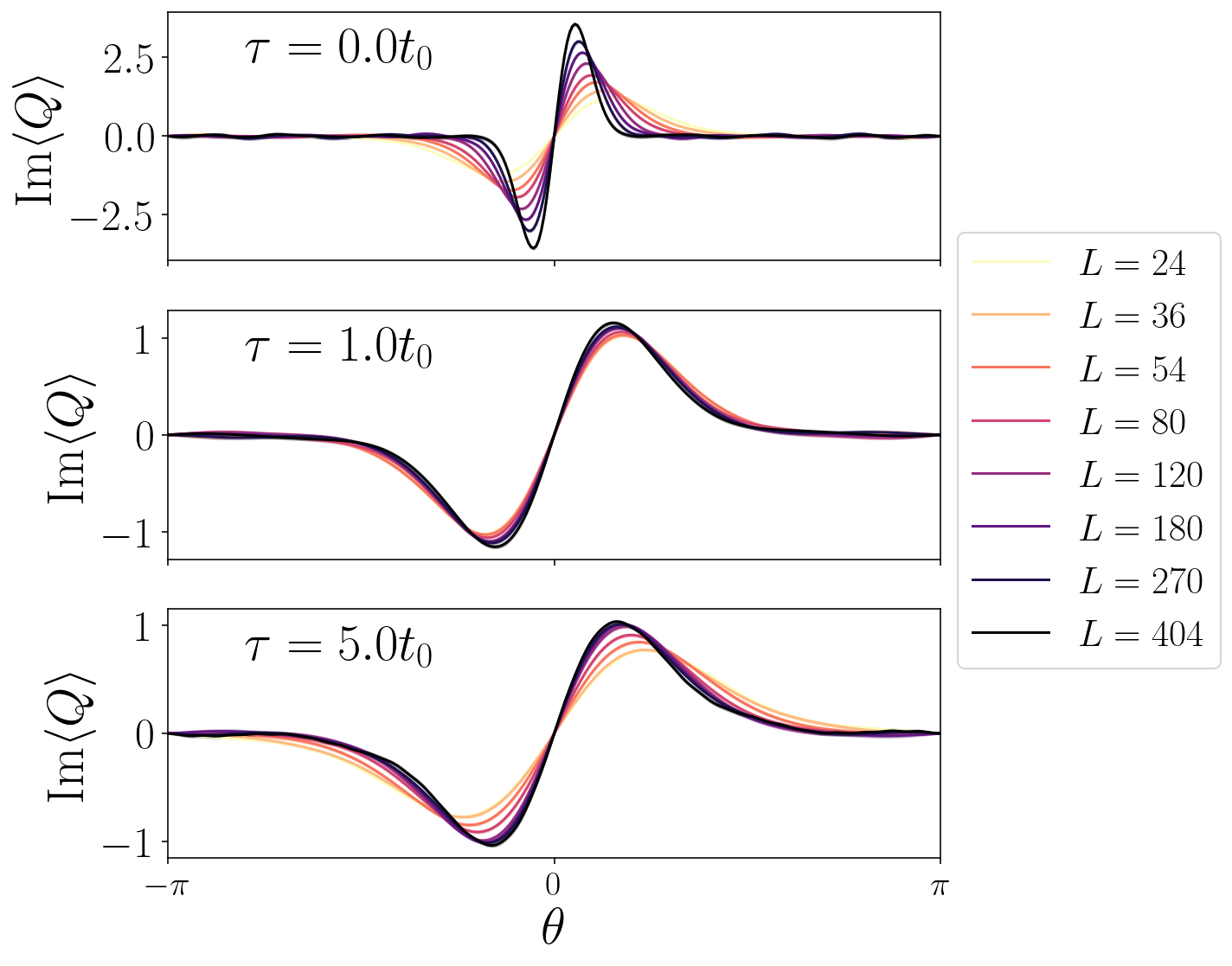}
      \caption{\label{fig:theta} Imaginary part of $\langle Q \rangle$ as a function of $\theta$. Simulation run with 10,000 measurements, measurements very 50 sweeps, 1,000 sweep thermalization. Note the different scaling of the $y$-axis.}
\end{figure}

We calculate the imaginary part of $\langle Q \rangle$ for arbitrary $\theta$ and plot the results for three values of the flow time $\tau$ in Fig~\ref{fig:theta}.
These plots demonstrate the divergence of the continuum limit in the $\tau=0$ and the flowed regimes. In the $\tau=0$ case, the slope increases sharply, reflecting the rapid divergence of $\chi_t$. However in the flowed regime, this divergence is much slower, reflecting the decreased values of $\chi_t$ shown in Fig.~\ref{fig:bietenholz}.

\section{Summary}

The topological behavior of the non-linear sigma model has been a topic of debate for several decades. Berg and L\"uscher \cite{berg1981} originally highlighted the discrepancy between the renormalization group hypothesis, that the topological susceptibility is zero in the continuum limit, and the numerical results of nonperturbative calculations, supported by a semi-classical instanton picture, which indicate that the topological susceptibility is ill-defined in the continuum limit. They suggested three possible causes:
\vspace*{-0.5\baselineskip}
\begin{enumerate}[itemsep=-0.5em]
    \item The definition of the topological charge does not scale to the continuum.
    \item There are ultraviolet divergences.
    \item There is no reasonable continuum limit.
\end{enumerate}

Our results at finite gradient flow time, for both $\theta = 0$ and at non-zero $\theta$,  support the numerical results of \cite{bietenholz2018} at $\theta=0$ and indicate the presence of a divergence in the continuum limit. Pre-reflectively, this may suggest that ultraviolet divergences are not the origin of this divergence, because the flow suppresses ultraviolet fluctuations and introduces no new divergences in the renormalized NLSM \cite{makino2015a}. The story of this divergence is, however, more subtle and is consistent with divergences at finite flow time. The topological susceptibility (and all other even moments of the topological charge) are divergent in the continuum limit, while differences between cumulants of the topological charge are finite \cite{Nogradi:2012dj}. This indicates that there is a single, nonperturbatively generated divergence that can be removed by introducing a single counter-term in the Lagrangian, proportional to the identity. In the absence of any such counter-term, the NLSM at zero flow time is not completely renormalized, and introducing the flow cannot remove this divergence. One future application of the gradient flow to the study of the topology of the NLSM, could be, however, to define a nonperturbative renormalization scheme for this counter term, and study the corresponding running of its coupling.

\section{Acknowledgments}

\noindent We thank Daniel Nogradi for enlightening discussions on the NLSM at nonzero $\theta$ during this conference, for drawing our attention to \cite{Nogradi:2012dj}, and for reading an early version of the manuscript. CJM is supported in part by USDOE grant No.~DE-AC05-06OR23177, under which Jefferson Science Associates, LLC, manages and operates Jefferson Lab.

\setlength{\itemsep}{-2pt}
\bibliographystyle{JHEP}
\bibliography{library}
\end{document}